\documentclass[twoside,a4paper]{article}
\usepackage{amsmath}
\usepackage{amssymb,bbold}
\usepackage{fullpage,xspace}
\usepackage{gnuplottex} %
\usepackage{graphics}
\usepackage{listings}
\usepackage{xcolor,svgcolor}
\usepackage{todonotes}
\usepackage{booktabs}

\lstdefinestyle{slp}{
 basicstyle=\ttfamily\footnotesize,
 breaklines=true,
 breakatwhitespace=true,
 breakindent=0pt,
 string=[d]{;},
 keywordstyle=\color{blue},
 showstringspaces=false
}
\usepackage{ifpdf}
\ifpdf%
\usepackage{ae} %
\usepackage{aeguill}
\fi
\usepackage[pdftex,backref=section]{hyperref}
\hypersetup{%
colorlinks,
citecolor=darkgreen,
filecolor=black,
linkcolor=darkred,
urlcolor=blue
}
\renewcommand*{\backref}[1]{}
\renewcommand*{\backrefalt}[4]{%
 \ifcase#1%
 \or\mbox{[Cited \S~#2.]}%
 \else\mbox{[Cited \S~#2.]}%
 \fi%
}
\makeatletter
\def\mathcolor#1#{\@mathcolor{#1}}
\def\@mathcolor#1#2#3{%
 \protect\leavevmode
 \begingroup
 \color#1{#2}#3%
 \endgroup
}
\makeatother
\usepackage[nameinlink,capitalize,noabbrev]{cleveref}
\Crefname{proposition}{Proposition}{Propositions}
\crefname{proposition}{Proposition}{Propositions}
\crefname{equation}{Eq.}{Eqs.} %
\Crefname{equation}{Equation}{Equations}
\crefformat{footnote}{#2\footnotemark[#1]#3}
\newcommand{\row}[2]{#1_{#2,*}}

\newcommand{\pnorm}[1]{\left\|#1\right\|_p}
\newcommand{\psnorm}[1]{\left\|#1\right\|_{p^\star}}
\newcommand{\qnorm}[1]{\left\|#1\right\|_q}
\newcommand{\qsnorm}[1]{\left\|#1\right\|_{q^\star}}

\newcommand{\ulp}{\varepsilon}

\newcommand{\GF}[2]{\ensuremath{\gamma_{#1,#2}}} %
\newcommand{\bigO}[1]{\ensuremath{O\mathopen{}\left({#1}\right)\mathclose{}}}
\newcommand{\bbigO}[1]{\ensuremath{O\!\bigl({#1}\bigl)}}

\newcommand{\firstdim}{\ensuremath{m}}
\newcommand{\seconddim}{\ensuremath{k}}
\newcommand{\thirddim}{\ensuremath{n}}
\newcommand{\Matrices}[2]{\ensuremath{{{#1}^{#2}}}}
\newcommand{\LRPRepresentation}[3]{{\ensuremath{\left[{{#1};{#2};{#3}}\right]}}}
\newcommand{\HadamardProduct}[2]{\ensuremath{{#1}\odot{#2}}}

\newcommand{\SLP}{{\textsc{slp}}\xspace}
\newcommand{\SLPs}{{\textsc{slp}s}\xspace}

\newcommand{\vectorization}[1]{\ensuremath{\textup{Vect}{(#1)}}}%
\newcommand{\matrixsize}[2]{\ensuremath{{{#1}{\times}{#2}}}}
\newcommand{\mat}[1]{\ensuremath{\mathsf{#1}}}%
\newcommand{\MatrixProduct}[2]{\ensuremath{{{\mat{#1}}\cdot{\mat{#2}}}}}

\newenvironment{smatrix}{\left[\begin{smallmatrix}}{\end{smallmatrix}\right]}
\newcommand{\Transpose}[1]{\ensuremath{{{#1}}^{\intercal}}}

\newcommand{\RR}{\ensuremath{\mathbb{R}}}
\newcommand{\ZZ}{\ensuremath{\mathbb{Z}}}
\newcommand{\FMMA}[4]{\ensuremath{{\langle{{#1}{\times}{#2}{\times}{#3}{:}{#4}}\rangle}}}
\newcommand{\plinoptdata}[1]{\href{https://github.com/jgdumas/plinopt/tree/main/data}{\url{#1}}}
\newtheorem{theorem}{Theorem}[section]

\newtheorem{remark}{Remark}[section]

\newcommand{\plinopt}{\href{https://github.com/jgdumas/plinopt}{\textsc{PLinOpt}}}
\title{A more accurate rational non-commutative algorithm for multiplying~\(\matrixsize{4}{4}\) matrices using~\(48\) multiplications}
\author{%
 Jean-Guillaume Dumas, \\
 \href{mailto:Jean-Guillaume.Dumas@univ-grenoble-alpes.fr}{\texttt{Jean-Guillaume.Dumas@univ-grenoble-alpes.fr}}\\
 Univ.\ Grenoble Alpes, \textsc{cnrs}, Grenoble \textsc{inp} -- \textsc{uga}, \textsc{ljk},\\
 38000 Grenoble, France
 \and
 Cl\'ement Pernet, \\
 \href{mailto:clement.pernet@univ-grenoble-alpes.fr}{\texttt{Clement.Pernet@univ-grenoble-alpes.fr}} \\
 Univ.\ Grenoble Alpes, \textsc{cnrs}, Grenoble \textsc{inp} -- \textsc{uga}, \textsc{ljk},\\
 38000 Grenoble, France
 \and
 Alexandre Sedoglavic,\\
 \href{mailto:Alexandre.Sedoglavic@univ-lille.fr}{\texttt{Alexandre.Sedoglavic@univ-lille.fr}}\\
 Univ.\ Lille, \textsc{cnrs}, \textsc{umr} 9189 \textsc{cris}t\textsc{al},\\
 F-59000 Lille, France
}
\begin{document}
\maketitle
\begin{abstract}
We propose a \emph{more accurate} variant of an algorithm for multiplying~\(\matrixsize{4}{4}\) matrices using~\(48\) multiplications over any ring containing an inverse of~\(2\).
This algorithm achieves an error bound exponent of only~\(\log_{4}\!\gamma_{\infty,2}\approx{2.335}\).
In practice, it also reaches a better accuracy w.r.t.\ max-norm, when compared to previously known such fast algorithms.
Furthermore, we propose a straight line program of this algorithm, giving a leading constant in its complexity bound of~\({\frac{316}{32}n^{2+\log_{4}\!{3}}+o\!\left({n^{2+\log_{4}\!{3}}}\right)}\) operations over any ring containing an inverse of~\(2\).
\end{abstract}
\section{Introduction}
An algorithm to multiply two~\(\matrixsize{4}{4}\) complex-valued matrices requiring only~\(48\) non-commutative multiplications was introduced in~\cite{alphaevolve}\footnote{A previous similar result was also announced in~\cite{Kaporin:2024ab} but this tensor decomposition could not be expressed without complex number as shown in~\cite[\S~1.2]{Moran:2026aa}.}\! using a pipeline of large language models orchestrated by an evolutionary coding agent.
A matrix multiplication algorithm with that many non-commutative multiplications is denoted by~\FMMA{4}{4}{4}{48} in the sequel.
\par
An \emph{equivalent variant} of the associated \emph{tensor decomposition} defining this algorithm, but over the rationals (more precisely over any ring containing an inverse of~\(2\)), was then subsequently given in~\cite{Dumas:2025aa}.
\par
Most error analysis of sub-cubic time matrix multiplication algorithms~\cite{Brent:1970,demmel:2007a,bini:1980,BBDLS16,STVW26} are given in the max-norm setting: bounding the largest output error as a function of the max-norm product of the vectors of input matrix coefficients.
In this setting, Strassen's algorithm has shown the best accuracy bound, (proven minimal under some assumptions in~\cite{bini:1980}).
\par
In~\cite{jgd:2024:accurate,Dumas:2026:autoaccurate}, the authors relaxed this setting by shifting the focus to the~\(2\)-norm for input and/or output; that allowed them to propose a~\FMMA{2}{2}{2}{7} variant with an improved accuracy bound.
Experiments show that this variant performs best even when measuring
the max-norm of the error bound.
\par
We present in this note a variant of the recent~\FMMA{4}{4}{4}{48}
algorithm over the rationals (again in the same orbit under De~Groot
\emph{isotropies}~\cite{groot:1978a}) that is more numerically
accurate w.r.t.\ max-norm in practice.
In particular, our new variant improves on the error bound exponent,
from~\({\log_{2}\!\gamma_{\infty,2}\approx{2.577}}\)
(resp.~\({\log_{4}\!\gamma_{\infty,2}\approx{2.628}}\)) for the
algorithm of~\cite{jgd:2024:accurate} (resp.~\cite{Dumas:2025aa}), to
now only~\({\log_{4}\!\gamma_{\infty,2}\approx{2.335}}\)
(see~\cref{tab:gfcomp} for more details).
\par
\Cref{ssec:newalgo} presents our new variant using a \textsc{lpr}
representation associated to matrix multiplication tensor
decomposition described in \Cref{sec:LRPdef}.
A detailed presentation of the framework used in this note could be
found in~\cite{Dumas:2025aa,Dumas:2026:autoaccurate} (see~\cite{Landsberg:2016ab}
for a more detailed reference).
\Cref{ssec:theoreticalbounds} is devoted to the comparison between the
error bound of this new algorithm with most common other fast ones.
\par
Then, \cref{sec:NumericalSchemeAndComplexity} shows how to implement
the associated theoretical algorithm and makes explicit the associated
complexity
bound:~\({\frac{316}{32}n^{2+\log_{4}\!{3}}-\frac{284}{32}n^{2}\approx{9.875n^{2.79248}}}\).
The cost for a better accuracy is thus a slight increase of the
operations cost when compared to the algorithms introduced
in~\cite{Dumas:2025aa}
(namely~\({{\frac{262}{32}n^{2+\log_{4}\!{3}}-\frac{230}{32}n^{2}}\approx{8.1875n^{2.79248}}}\)).
We also give in~\cref{sec:AlternativeBasis} an alternative bases
variant with similar accuracy.
Its complexity bound is
only~\({7n^{2+\log_{4}\!{3}}+o\!\left({n^{2+\log_{4}\!{3}}}\right)}\),
exactly matching that of the alternative bases variant
of~\cite{Dumas:2025aa}.
\par
Finally, \cref{sec:bench} presents an associated accuracy benchmark on
randomly sampled floating point matrices.

\section{A new variant for the~\({{4}\times{4}}\) by~\({{4}\times{4}}\) matrix multiplication with error bound exponent~\(2.335\)}\label{sec:TheAlgorithm}
Given an~\(\matrixsize{\firstdim}{\seconddim}\) matrix~\(\mat{A}\), we denote by~\(\row{\mat{A}}{i}\) the~\(i\)th row and by~\(\vectorization{\mat{A}}\) the row-major vectorization of this matrix, i.e.\ the vector~\(v\) in~\(\RR^{\firstdim\seconddim}\) of the matrix coefficients such that~\({v_{i\seconddim+j} = a_{i,j}}\).
\subsection{LRP representation of a matrix multiplication algorithm}\label{sec:LRPdef}
As for any bilinear operator, a matrix multiplication algorithm can be represented as a triple of matrices~\({\mat{L},\mat{R},\mat{P}}\) as follows:
\begin{equation}\label{eq:HMRepresentation2MatrixMultiplicationFormula}
\begin{array}{cccc}
\beta: & {{\Matrices{\RR}{\matrixsize{\firstdim}{\seconddim}}}\times{\Matrices{\RR}{\matrixsize{\seconddim}{\thirddim}}}}
&\rightarrow&
\Matrices{\RR}{\matrixsize{\firstdim\thirddim}{1}},\\
&(\mat{A},\mat{B}) & \mapsto & \MatrixProduct%
{\mat{P}}
{\left( \HadamardProduct
 {\left(\MatrixProduct{L}{\vectorization{\mat{A}}}\right)}%
 {\left(\MatrixProduct{R}{\vectorization{\mat{B}}}\right)}%
\right)},
\end{array}
\end{equation}
where~\(\HadamardProduct{}{}\) stands for the Hadamard product.
For instance the standard~\FMMA{2}{2}{2}{8} algorithm for a~\(\matrixsize{2}{2}\) matrix product~\({
\begin{smatrix} c_{11} & c_{12}\\c_{21}&c_{22}\end{smatrix}=
\begin{smatrix} a_{11} & a_{12}\\a_{21}&a_{22}\end{smatrix}\cdot
	\begin{smatrix} b_{11} & b_{12}\\b_{21}&b_{22}\end{smatrix}}\),
can be represented as:
\begin{equation}
\begin{bmatrix}c_{11} \\ c_{12} \\ c_{21} \\ c_{22}\end{bmatrix}=
\begin{bmatrix}
1&1&0&0&0&0&0&0\\
0&0&1&1&0&0&0&0\\
0&0&0&0&1&1&0&0\\
0&0&0&0&0&0&1&1\\
\end{bmatrix}
\left(
\begin{bmatrix}
1&0&0&0\\
0&1&0&0\\
1&0&0&0\\
0&1&0&0\\
0&0&1&0\\
0&0&0&1\\
0&0&1&0\\
0&0&0&1\\
\end{bmatrix}
\begin{bmatrix}a_{11} \\ a_{12} \\ a_{21} \\ a_{22}\end{bmatrix}\right)
\HadamardProduct{}{}
\left(\begin{bmatrix}
1&0&0&0\\
0&0&1&0\\
0&1&0&0\\
0&0&0&1\\
1&0&0&0\\
0&0&1&0\\
0&1&0&0\\
0&0&0&1\\
\end{bmatrix}
\begin{bmatrix}b_{11} \\ b_{12} \\ b_{21} \\ b_{22}\end{bmatrix}\right).
\end{equation}

\subsection{The new algorithmic variant}\label{ssec:newalgo}
We propose a new variant of the \FMMA{4}{4}{4}{48} matrix
multiplication algorithm defined by the \textsc{lrp} representation
given in~\cref{alg:444accurate}.
To find this algorithm, we considered the De~Groot isotropy orbit
(see~\cite{groot:1978a} or, e.g., the presentation
in~\cite[Lemma~8]{Dumas:2026:autoaccurate}) of the rational
\FMMA{4}{4}{4}{48} algorithm
of~\cite{Dumas:2025aa}:
\begin{enumerate}
\item As in~\cite[\S~4]{Dumas:2026:autoaccurate} we first compute the
  minimal $\gamma_2$ growth factor along this orbit,
  see~\cref{rem:gamma2};
\item From this point, we then look for {\emph{orthogonal}}
    isotropies that produce sparser \textsc{lrp} representations.
\end{enumerate}
Indeed, restricting to a sub-orbit with only orthogonal isotropies
will not change the $2$-norm, and will therefore preserve the
$\gamma_2$ growth factor.
In practice, we performed some random walks with sparse orthogonal
Householder rank-$1$ updates ($I-2\vec{u}\Transpose{\vec{u}}$, with a
unit vector $\vec{u}$) using \plinopt's \texttt{orbiter}.
Then, even though sparser \textsc{lrp} representations do not always produce
the least number of operations (see for instance the \FMMA{4}{4}{4}{7}
variants of Strassen or Winograd), the one we obtained has
quite efficient \SLPs, see~\cref{sec:NumericalSchemeAndComplexity}.

\begin{table}[htbp]\centering
\scalebox{.8}{%
\(\begin{smatrix}
0&2&0&2&0&-2&0&-2&0&0&0&0&0&0&0&0\\
-2&0&-2&0&-2&0&-2&0&0&0&0&0&0&0&0&0\\
0&0&0&0&-1&-1&0&0&0&0&0&0&1&1&0&0\\
0&0&0&0&0&0&0&0&2&0&-2&0&-2&0&2&0\\
-1&0&-1&0&0&0&0&0&1&0&1&0&0&0&0&0\\
0&0&0&0&1&1&0&0&0&0&0&0&1&1&0&0\\
0&0&0&0&0&0&-1&-1&0&0&0&0&0&0&1&1\\
0&0&-1&1&0&0&0&0&0&0&-1&1&0&0&0&0\\
0&0&0&0&1&0&1&0&0&0&0&0&1&0&1&0\\
1&0&-1&0&0&0&0&0&1&0&-1&0&0&0&0&0\\
0&0&0&0&0&0&0&0&2&-2&0&0&2&-2&0&0\\
0&0&1&1&0&0&0&0&0&0&1&1&0&0&0&0\\
-1&1&0&0&0&0&0&0&1&-1&0&0&0&0&0&0\\
0&0&0&0&-1&1&0&0&0&0&0&0&-1&1&0&0\\
2&0&2&0&-2&0&-2&0&0&0&0&0&0&0&0&0\\
-1&0&1&0&0&0&0&0&1&0&-1&0&0&0&0&0\\
-2&0&2&0&-2&0&2&0&0&0&0&0&0&0&0&0\\
0&0&-2&2&0&0&2&-2&0&0&0&0&0&0&0&0\\
0&0&0&0&0&0&0&0&0&2&0&-2&0&-2&0&2\\
2&2&0&0&2&2&0&0&0&0&0&0&0&0&0&0\\
0&0&0&0&0&0&0&0&0&2&0&-2&0&2&0&-2\\
0&0&2&2&0&0&2&2&0&0&0&0&0&0&0&0\\
0&0&0&0&0&0&0&0&2&2&0&0&-2&-2&0&0\\
0&0&0&0&0&1&0&-1&0&0&0&0&0&-1&0&1\\
0&0&-1&1&0&0&0&0&0&0&1&-1&0&0&0&0\\
0&0&0&0&0&1&0&1&0&0&0&0&0&1&0&1\\
0&1&0&-1&0&0&0&0&0&1&0&-1&0&0&0&0\\
0&0&2&2&0&0&-2&-2&0&0&0&0&0&0&0&0\\
0&0&0&0&0&1&0&1&0&0&0&0&0&-1&0&-1\\
0&0&0&0&0&0&0&0&0&2&0&2&0&2&0&2\\
0&0&0&0&0&0&0&0&0&0&-2&2&0&0&-2&2\\
0&0&0&0&0&0&0&0&2&-2&0&0&-2&2&0&0\\
1&0&0&-1&-1&0&0&1&-1&2&0&-1&-1&2&0&-1\\
-2&0&2&0&-2&-2&0&0&0&0&0&0&0&0&0&0\\
0&0&1&1&0&0&0&0&0&-1&0&1&0&0&0&0\\
-1&0&0&1&-1&-2&0&-1&-1&0&0&1&1&2&0&1\\
-1&0&0&1&-1&0&0&1&1&-2&0&1&-1&2&0&-1\\
-1&0&-1&0&0&0&0&0&1&-1&0&0&0&0&0&0\\
0&0&0&0&0&0&-1&-1&0&0&0&0&0&-1&0&1\\
0&0&0&0&0&0&0&0&2&0&-2&0&-2&-2&0&0\\
0&0&-2&2&0&-2&0&-2&0&0&0&0&0&0&0&0\\
0&0&0&0&-1&0&-1&0&0&0&0&0&-1&1&0&0\\
-1&0&-2&-1&1&0&2&1&1&0&0&-1&1&0&0&-1\\
0&0&0&0&0&0&0&0&0&0&-2&2&0&2&0&2\\
-1&0&2&-1&-1&0&0&1&1&0&-2&1&-1&0&0&1\\
-1&0&-2&-1&-1&0&-2&-1&1&0&0&-1&-1&0&0&1\\
-1&0&2&-1&-1&0&0&1&-1&0&2&-1&1&0&0&-1\\
-1&0&0&1&-1&-2&0&-1&1&0&0&-1&-1&-2&0&-1\\
\end{smatrix}\quad\quad
\begin{smatrix}
0&0&0&0&0&1&0&1&0&0&0&0&0&0&0&0\\
0&\frac{1}{2}&0&-\frac{1}{2}&0&-\frac{1}{2}&0&\frac{1}{2}&0&\frac{1}{2}&0&-\frac{1}{2}&0&-\frac{1}{2}&0&\frac{1}{2}\\
0&1&-1&0&0&0&0&0&0&0&0&0&0&0&0&0\\
0&0&0&0&0&0&0&0&1&0&1&0&0&0&0&0\\
-\frac{1}{2}&0&0&\frac{1}{2}&-\frac{1}{2}&0&0&\frac{1}{2}&-\frac{1}{2}&0&0&\frac{1}{2}&-\frac{1}{2}&0&0&\frac{1}{2}\\
-\frac{1}{2}&0&0&-\frac{1}{2}&-\frac{1}{2}&0&0&-\frac{1}{2}&\frac{1}{2}&0&0&\frac{1}{2}&\frac{1}{2}&0&0&\frac{1}{2}\\
0&0&0&0&0&0&0&0&0&-1&1&0&0&0&0&0\\
0&-\frac{1}{2}&-\frac{1}{2}&0&0&\frac{1}{2}&\frac{1}{2}&0&0&-\frac{1}{2}&-\frac{1}{2}&0&0&\frac{1}{2}&\frac{1}{2}&0\\
0&0&0&0&0&0&0&0&1&0&0&1&0&0&0&0\\
0&1&1&0&0&0&0&0&0&0&0&0&0&0&0&0\\
1&0&-1&0&0&0&0&0&0&0&0&0&0&0&0&0\\
0&-\frac{1}{2}&-\frac{1}{2}&0&0&-\frac{1}{2}&-\frac{1}{2}&0&0&-\frac{1}{2}&-\frac{1}{2}&0&0&-\frac{1}{2}&-\frac{1}{2}&0\\
0&0&0&0&-1&0&0&1&0&0&0&0&0&0&0&0\\
\frac{1}{2}&0&0&\frac{1}{2}&-\frac{1}{2}&0&0&-\frac{1}{2}&-\frac{1}{2}&0&0&-\frac{1}{2}&\frac{1}{2}&0&0&\frac{1}{2}\\
0&1&0&1&0&0&0&0&0&0&0&0&0&0&0&0\\
-\frac{1}{2}&0&0&\frac{1}{2}&-\frac{1}{2}&0&0&\frac{1}{2}&\frac{1}{2}&0&0&-\frac{1}{2}&\frac{1}{2}&0&0&-\frac{1}{2}\\
0&-\frac{1}{2}&0&\frac{1}{2}&0&\frac{1}{2}&0&-\frac{1}{2}&0&\frac{1}{2}&0&-\frac{1}{2}&0&-\frac{1}{2}&0&\frac{1}{2}\\
0&\frac{1}{2}&0&\frac{1}{2}&0&-\frac{1}{2}&0&-\frac{1}{2}&0&-\frac{1}{2}&0&-\frac{1}{2}&0&\frac{1}{2}&0&\frac{1}{2}\\
0&0&0&0&0&0&0&0&0&0&0&0&1&0&1&0\\
0&0&0&0&0&1&0&-1&0&0&0&0&0&0&0&0\\
-\frac{1}{2}&0&\frac{1}{2}&0&-\frac{1}{2}&0&\frac{1}{2}&0&\frac{1}{2}&0&-\frac{1}{2}&0&\frac{1}{2}&0&-\frac{1}{2}&0\\
0&0&0&0&0&0&0&0&0&0&0&0&0&-1&0&1\\
\frac{1}{2}&0&\frac{1}{2}&0&\frac{1}{2}&0&\frac{1}{2}&0&\frac{1}{2}&0&\frac{1}{2}&0&\frac{1}{2}&0&\frac{1}{2}&0\\
0&\frac{1}{2}&-\frac{1}{2}&0&0&-\frac{1}{2}&\frac{1}{2}&0&0&-\frac{1}{2}&\frac{1}{2}&0&0&\frac{1}{2}&-\frac{1}{2}&0\\
0&0&0&0&0&0&0&0&0&0&0&0&-1&0&0&1\\
0&0&0&0&0&0&0&0&0&0&0&0&1&0&0&1\\
0&0&0&0&0&1&1&0&0&0&0&0&0&0&0&0\\
0&\frac{1}{2}&0&\frac{1}{2}&0&\frac{1}{2}&0&\frac{1}{2}&0&-\frac{1}{2}&0&-\frac{1}{2}&0&-\frac{1}{2}&0&-\frac{1}{2}\\
0&\frac{1}{2}&-\frac{1}{2}&0&0&-\frac{1}{2}&\frac{1}{2}&0&0&\frac{1}{2}&-\frac{1}{2}&0&0&-\frac{1}{2}&\frac{1}{2}&0\\
\frac{1}{2}&0&-\frac{1}{2}&0&\frac{1}{2}&0&-\frac{1}{2}&0&\frac{1}{2}&0&-\frac{1}{2}&0&\frac{1}{2}&0&-\frac{1}{2}&0\\
0&0&0&0&0&0&0&0&1&0&-1&0&0&0&0&0\\
-\frac{1}{2}&0&-\frac{1}{2}&0&\frac{1}{2}&0&\frac{1}{2}&0&-\frac{1}{2}&0&-\frac{1}{2}&0&\frac{1}{2}&0&\frac{1}{2}&0\\
\frac{1}{2}&0&-1&0&\frac{1}{2}&0&0&0&-\frac{1}{2}&0&0&0&-\frac{1}{2}&0&0&0\\
\frac{1}{4}&-\frac{1}{4}&\frac{1}{4}&\frac{1}{4}&\frac{1}{4}&\frac{3}{4}&\frac{1}{4}&-\frac{3}{4}&-\frac{1}{4}&\frac{1}{4}&-\frac{1}{4}&-\frac{1}{4}&\frac{1}{4}&-\frac{1}{4}&\frac{1}{4}&\frac{1}{4}\\
0&0&\frac{1}{2}&0&0&1&\frac{1}{2}&0&0&0&\frac{1}{2}&0&0&0&\frac{1}{2}&0\\
\frac{1}{4}&\frac{3}{4}&-\frac{3}{4}&\frac{1}{4}&\frac{1}{4}&-\frac{1}{4}&\frac{1}{4}&\frac{1}{4}&-\frac{1}{4}&\frac{1}{4}&-\frac{1}{4}&-\frac{1}{4}&\frac{1}{4}&-\frac{1}{4}&\frac{1}{4}&\frac{1}{4}\\
0&0&-\frac{1}{2}&0&0&0&\frac{1}{2}&0&0&0&-\frac{1}{2}&0&1&0&\frac{1}{2}&0\\
\frac{1}{2}&0&0&0&\frac{1}{2}&0&0&-1&\frac{1}{2}&0&0&0&\frac{1}{2}&0&0&0\\
0&-\frac{1}{2}&0&0&0&\frac{1}{2}&0&0&0&\frac{1}{2}&-1&0&0&-\frac{1}{2}&0&0\\
-\frac{1}{4}&-\frac{1}{4}&-\frac{1}{4}&\frac{1}{4}&-\frac{1}{4}&-\frac{1}{4}&-\frac{1}{4}&\frac{1}{4}&-\frac{3}{4}&\frac{1}{4}&-\frac{3}{4}&-\frac{1}{4}&-\frac{1}{4}&-\frac{1}{4}&-\frac{1}{4}&\frac{1}{4}\\
-\frac{1}{4}&-\frac{1}{4}&\frac{1}{4}&-\frac{1}{4}&-\frac{1}{4}&\frac{3}{4}&\frac{1}{4}&\frac{3}{4}&\frac{1}{4}&\frac{1}{4}&-\frac{1}{4}&\frac{1}{4}&-\frac{1}{4}&-\frac{1}{4}&\frac{1}{4}&-\frac{1}{4}\\
0&0&0&\frac{1}{2}&0&0&0&-\frac{1}{2}&-1&0&0&-\frac{1}{2}&0&0&0&\frac{1}{2}\\
0&1&0&\frac{1}{2}&0&0&0&\frac{1}{2}&0&0&0&-\frac{1}{2}&0&0&0&-\frac{1}{2}\\
-\frac{1}{4}&\frac{1}{4}&\frac{1}{4}&\frac{1}{4}&-\frac{1}{4}&\frac{1}{4}&\frac{1}{4}&\frac{1}{4}&-\frac{3}{4}&-\frac{1}{4}&\frac{3}{4}&-\frac{1}{4}&-\frac{1}{4}&\frac{1}{4}&\frac{1}{4}&\frac{1}{4}\\
\frac{1}{4}&-\frac{1}{4}&-\frac{1}{4}&-\frac{1}{4}&\frac{1}{4}&-\frac{1}{4}&-\frac{1}{4}&-\frac{1}{4}&-\frac{1}{4}&\frac{1}{4}&\frac{1}{4}&\frac{1}{4}&-\frac{3}{4}&-\frac{1}{4}&-\frac{1}{4}&\frac{3}{4}\\
0&\frac{1}{2}&0&0&0&-\frac{1}{2}&0&0&0&\frac{1}{2}&0&0&0&-\frac{1}{2}&0&1\\
-\frac{1}{4}&\frac{3}{4}&\frac{3}{4}&\frac{1}{4}&-\frac{1}{4}&-\frac{1}{4}&-\frac{1}{4}&\frac{1}{4}&\frac{1}{4}&\frac{1}{4}&\frac{1}{4}&-\frac{1}{4}&-\frac{1}{4}&-\frac{1}{4}&-\frac{1}{4}&\frac{1}{4}\\
\frac{1}{4}&\frac{1}{4}&-\frac{1}{4}&\frac{1}{4}&\frac{1}{4}&\frac{1}{4}&-\frac{1}{4}&\frac{1}{4}&-\frac{1}{4}&-\frac{1}{4}&\frac{1}{4}&-\frac{1}{4}&-\frac{3}{4}&\frac{1}{4}&-\frac{1}{4}&-\frac{3}{4}\\
\end{smatrix}\)
}
\smallskip
\par\noindent
\scalebox{0.7}{%
\(\begin{smatrix}
\frac{1}{4}&0&-\frac{1}{4}&0&0&\frac{1}{4}&0&-\frac{1}{4}&0&\frac{1}{4}&\frac{1}{4}&0&-1&0&0&\frac{1}{4}&0&0&0&-\frac{1}{4}&0&0&0&0&-\frac{1}{4}&\frac{1}{4}&0&0&-\frac{1}{4}&0&0&-\frac{1}{4}&\frac{1}{4}&-\frac{1}{4}&0&0&\frac{1}{4}&-1&0&0&-\frac{1}{4}&0&0&0&0&0&\frac{1}{4}&-\frac{1}{4}\\
\frac{1}{4}&0&\frac{1}{4}&0&0&-\frac{1}{4}&0&\frac{1}{4}&0&\frac{1}{4}&0&0&0&0&0&\frac{1}{4}&0&0&0&\frac{1}{4}&0&-\frac{1}{4}&0&0&\frac{1}{4}&\frac{1}{4}&0&-\frac{1}{4}&-\frac{1}{4}&0&0&0&0&\frac{1}{4}&0&-\frac{1}{4}&0&0&0&0&-\frac{1}{4}&0&-\frac{1}{4}&0&\frac{1}{4}&-\frac{1}{4}&0&0\\
-\frac{1}{4}&0&-\frac{1}{4}&0&0&\frac{1}{4}&0&\frac{1}{4}&0&\frac{1}{4}&0&0&0&0&0&\frac{1}{4}&0&0&\frac{1}{4}&-\frac{1}{4}&-\frac{1}{4}&0&0&0&\frac{1}{4}&-\frac{1}{4}&1&0&\frac{1}{4}&0&0&0&-\frac{1}{4}&-\frac{1}{4}&1&\frac{1}{4}&\frac{1}{4}&0&0&0&\frac{1}{4}&0&0&0&\frac{1}{4}&0&0&0\\
\frac{1}{4}&\frac{1}{4}&-\frac{1}{4}&0&0&\frac{1}{4}&0&\frac{1}{4}&0&-\frac{1}{4}&0&0&0&0&\frac{1}{4}&-\frac{1}{4}&0&0&0&-\frac{1}{4}&0&0&0&0&\frac{1}{4}&\frac{1}{4}&0&0&-\frac{1}{4}&0&0&0&0&-\frac{1}{4}&0&0&0&0&0&0&-\frac{1}{4}&0&\frac{1}{4}&0&0&-\frac{1}{4}&-\frac{1}{4}&-\frac{1}{4}\\
0&0&\frac{1}{4}&0&0&-\frac{1}{4}&0&-\frac{1}{4}&0&-\frac{1}{4}&-\frac{1}{4}&0&0&-1&0&-\frac{1}{4}&\frac{1}{4}&\frac{1}{4}&0&0&0&0&0&0&-\frac{1}{4}&\frac{1}{4}&0&0&-\frac{1}{4}&0&0&-\frac{1}{4}&-\frac{1}{4}&-\frac{1}{4}&0&-\frac{1}{4}&\frac{1}{4}&0&0&0&\frac{1}{4}&1&0&0&-\frac{1}{4}&0&0&0\\
0&0&-\frac{1}{4}&0&0&\frac{1}{4}&0&\frac{1}{4}&0&-\frac{1}{4}&0&0&0&0&0&-\frac{1}{4}&\frac{1}{4}&-\frac{1}{4}&0&0&0&-\frac{1}{4}&0&0&\frac{1}{4}&\frac{1}{4}&0&\frac{1}{4}&-\frac{1}{4}&0&0&0&0&-\frac{1}{4}&0&0&0&0&0&0&-\frac{1}{4}&0&\frac{1}{4}&0&0&-\frac{1}{4}&-\frac{1}{4}&-\frac{1}{4}\\
0&0&\frac{1}{4}&0&0&-\frac{1}{4}&0&\frac{1}{4}&0&-\frac{1}{4}&0&0&0&0&0&-\frac{1}{4}&\frac{1}{4}&-\frac{1}{4}&\frac{1}{4}&0&\frac{1}{4}&0&0&1&\frac{1}{4}&-\frac{1}{4}&0&0&\frac{1}{4}&0&0&0&\frac{1}{4}&-\frac{1}{4}&0&0&\frac{1}{4}&0&1&0&-\frac{1}{4}&0&0&0&0&0&-\frac{1}{4}&\frac{1}{4}\\
0&\frac{1}{4}&\frac{1}{4}&0&0&-\frac{1}{4}&0&\frac{1}{4}&0&\frac{1}{4}&0&0&0&0&-\frac{1}{4}&\frac{1}{4}&-\frac{1}{4}&-\frac{1}{4}&0&0&0&0&0&0&\frac{1}{4}&\frac{1}{4}&0&0&-\frac{1}{4}&0&0&0&0&\frac{1}{4}&0&-\frac{1}{4}&0&0&0&0&-\frac{1}{4}&0&-\frac{1}{4}&0&\frac{1}{4}&-\frac{1}{4}&0&0\\
0&0&-\frac{1}{4}&0&0&-\frac{1}{4}&0&-\frac{1}{4}&0&\frac{1}{4}&\frac{1}{4}&0&0&0&0&-\frac{1}{4}&0&0&0&0&0&0&\frac{1}{4}&0&\frac{1}{4}&-\frac{1}{4}&0&0&-\frac{1}{4}&\frac{1}{4}&0&-\frac{1}{4}&\frac{1}{4}&0&0&0&\frac{1}{4}&0&0&\frac{1}{4}&0&0&0&\frac{1}{4}&0&0&\frac{1}{4}&\frac{1}{4}\\
0&0&\frac{1}{4}&0&0&\frac{1}{4}&0&\frac{1}{4}&0&\frac{1}{4}&0&-1&0&0&0&-\frac{1}{4}&0&0&0&0&0&\frac{1}{4}&-\frac{1}{4}&0&-\frac{1}{4}&-\frac{1}{4}&0&\frac{1}{4}&-\frac{1}{4}&\frac{1}{4}&0&0&0&0&-1&-\frac{1}{4}&0&0&0&-\frac{1}{4}&0&0&\frac{1}{4}&\frac{1}{4}&-\frac{1}{4}&\frac{1}{4}&0&0\\
0&0&-\frac{1}{4}&0&0&-\frac{1}{4}&0&\frac{1}{4}&0&\frac{1}{4}&0&0&0&0&0&-\frac{1}{4}&0&0&-\frac{1}{4}&0&\frac{1}{4}&0&\frac{1}{4}&0&-\frac{1}{4}&\frac{1}{4}&0&0&\frac{1}{4}&-\frac{1}{4}&0&0&\frac{1}{4}&0&0&\frac{1}{4}&-\frac{1}{4}&0&0&\frac{1}{4}&0&0&0&-\frac{1}{4}&-\frac{1}{4}&0&0&0\\
0&\frac{1}{4}&-\frac{1}{4}&0&1&-\frac{1}{4}&0&\frac{1}{4}&0&-\frac{1}{4}&0&0&0&0&\frac{1}{4}&\frac{1}{4}&0&0&0&0&0&0&\frac{1}{4}&0&-\frac{1}{4}&-\frac{1}{4}&0&0&-\frac{1}{4}&\frac{1}{4}&0&0&0&0&0&0&0&1&0&\frac{1}{4}&0&0&\frac{1}{4}&\frac{1}{4}&0&-\frac{1}{4}&-\frac{1}{4}&\frac{1}{4}\\
0&0&-\frac{1}{4}&\frac{1}{4}&0&-\frac{1}{4}&0&\frac{1}{4}&0&\frac{1}{4}&\frac{1}{4}&0&0&0&0&-\frac{1}{4}&0&0&0&0&0&0&0&0&-\frac{1}{4}&\frac{1}{4}&0&0&\frac{1}{4}&0&-\frac{1}{4}&\frac{1}{4}&\frac{1}{4}&0&0&\frac{1}{4}&-\frac{1}{4}&0&0&\frac{1}{4}&0&0&0&-\frac{1}{4}&-\frac{1}{4}&0&0&0\\
0&0&\frac{1}{4}&\frac{1}{4}&0&\frac{1}{4}&-1&-\frac{1}{4}&0&\frac{1}{4}&0&0&0&0&0&-\frac{1}{4}&0&0&0&0&0&-\frac{1}{4}&0&0&\frac{1}{4}&\frac{1}{4}&0&\frac{1}{4}&\frac{1}{4}&0&\frac{1}{4}&0&0&0&0&0&0&0&-1&\frac{1}{4}&0&0&\frac{1}{4}&\frac{1}{4}&0&-\frac{1}{4}&\frac{1}{4}&-\frac{1}{4}\\
0&0&-\frac{1}{4}&\frac{1}{4}&0&-\frac{1}{4}&0&-\frac{1}{4}&0&\frac{1}{4}&0&0&0&0&0&-\frac{1}{4}&0&0&\frac{1}{4}&0&\frac{1}{4}&0&0&0&\frac{1}{4}&-\frac{1}{4}&0&0&-\frac{1}{4}&0&\frac{1}{4}&0&\frac{1}{4}&0&0&0&\frac{1}{4}&0&0&\frac{1}{4}&0&0&0&\frac{1}{4}&0&0&\frac{1}{4}&\frac{1}{4}\\
0&-\frac{1}{4}&-\frac{1}{4}&-\frac{1}{4}&0&-\frac{1}{4}&0&-\frac{1}{4}&1&-\frac{1}{4}&0&0&0&0&\frac{1}{4}&\frac{1}{4}&0&0&0&0&0&0&0&0&\frac{1}{4}&\frac{1}{4}&0&0&\frac{1}{4}&0&\frac{1}{4}&0&0&0&0&\frac{1}{4}&0&0&0&-\frac{1}{4}&0&-1&\frac{1}{4}&\frac{1}{4}&\frac{1}{4}&\frac{1}{4}&0&0\\
\end{smatrix}\)
}
\caption{\LRPRepresentation{L}{R}{P} matrices of the accurate \FMMA{4}{4}{4}{48} matrix multiplication algorithm}\label{alg:444accurate}
\end{table}
\subsection{Accuracy bounds}\label{ssec:theoreticalbounds}
\par
Consider the product of an~\({{M}\times{K}}\) matrix~\(\mat{A}\) by a~\({{K}\times{N}}\) matrix~\(\mat{B}\).
It is computed by a~\({\langle m,k,n\rangle}\) algorithm represented by the matrices~\({\mat{L},\mat{R},\mat{P}}\) applied recursively on~\(\ell\) recursive levels and the resulting~\({{m_{0}}\times{k_{0}}}\) by~\({{k_{0}}\times{n_{0}}}\) products are performed using an algorithm~\(\beta\).
Here~\({{M={m_{0}m^{\ell}}},{K={k_0k^{\ell}}}}\) and~\({n=n_0n^{\ell}}\).
\par
The accuracy bound below uses any (possibly different)~\(p\)-norms and~\(q\)-norms for its left-hand-side,~\(\pnorm{\cdot}\) and right-hand side,~\(\qnorm{\cdot}\).
The associated dual norms, are denoted by~\(\psnorm{\cdot}\) and~\(\qsnorm{\cdot}\) respectively.
Note that, these are vector norms, hence~\(\pnorm{\mat{A}}\) for matrix~\(\mat{A}\) in~\(\RR^{\matrixsize{m}{n}}\) denotes~\(\pnorm{\vectorization{\mat{A}}}\) and is the~\(p\)-norm of the~\(mn\) dimensional vector of its coefficients, and not a matrix norm.
\begin{theorem}[{\cite[Th. 18]{Dumas:2026:autoaccurate}}]
	The forward error in computing the approximate value~\(\widehat{\beta(\mat{A},\mat{B})}\) of the product~\({\mat{C}={\mat{A}\times\mat{B}}}\) satisfies:
\begin{equation}
	\pnorm{\widehat{\beta(\mat{A},\mat{B})}-\mat{C}} \leq f_{p,q}\qnorm{\mat{A}}\qnorm{\mat{B}}\ulp+\bbigO{\ulp^2}\quad\textup{where}\quad
	f_{p,q} = \bigO{\gamma_{p,q}^\ell} =
    \bigO{{\left(\frac{K}{k_0}\right)}^{\log_{k}{\gamma_{p,q}}}},
\end{equation}
and
\begin{equation}\label{eq:GFpq}
\GF{p}{q} =
\pnorm{{\left(\sum_{i=1}^{r}\qsnorm{\row{\mat{L}}{i}}\qsnorm{\row{\mat{R}}{i}} |\mat{p_{j,i}}|\right)}_j}.
\end{equation}
\end{theorem}
\begin{remark}
The parameter~\(\GF{p}{q}\) is called the~\({(p,q)}\)-growth factor and \(f_{p,q}\) is the error bound function of the algorithm defined by the representation \LRPRepresentation{L}{R}{P} with respect to the~\(p\)- and the~\(q\)-norms.
Most results on the accuracy of fast matrix multiplications
use~\({p=q=\infty}\)~\cite{Brent:1970,bini:1980,demmel:2007a,BBDLS16,Higham:2002,STVW26}.
Following our approach in~\cite{Dumas:2026:autoaccurate}, we focus on the smoother~\(2\)-norm, namely~\(\gamma_{2,2}\) and~\(\gamma_{\infty,2}\).
Using these norms, Strassen's algorithm no longer has the most accurate bound among all \FMMA{2}{2}{2}{7} algorithms.
The accuracy parameter to be optimized is a smoother function and more importantly, these norms seem to better reflect the max-norm accuracy behavior in practice of the considered algorithms (see~\Cref{sec:bench}).
\end{remark}
\Cref{tab:gfcomp} displays the growth factors and the corresponding
error bound functions for all combinations of~\(\infty-\)
and~\(2\)-norms of the above algorithm compared to Strassen's
algorithm~\cite{strassen:1969}, Winograd's
algorithm~\cite{winograd:1977:complexite}, the authors'
accurate~\FMMA{2}{2}{2}{7}~\cite{jgd:2024:accurate}, and the authors'
rational~\FMMA{4}{4}{4}{48} algorithm~\cite{Dumas:2025aa}
over~\(\ZZ[1/2]\).

\begin{table}[htbp]
 \begin{center}
 \small
 \begin{tabular}{l|cc|cc|cc|cc
 }
 \toprule
 \((p,q)\) & \multicolumn{2}{c|}{\((\infty,\infty)\)}& \multicolumn{2}{c|}{\((\infty,2)\)} & \multicolumn{2}{c|}{\((2,\infty)\)} & \multicolumn{2}{c}{\((2,2)\)}
 \\
  & \(\gamma_{p,q}\) & \(f_{p,q}\) & \(\gamma_{p,q}\) & \(f_{p,q}\) & \(\gamma_{p,q}\) & \(f_{p,q}\)& \(\gamma_{p,q}\) & \(f_{p,q}\)\\
 \midrule
 \FMMA{2}{2}{2}{7} Win77 & 18 & \bigO{n^{4.170}} & 8 & \bigO{n^3} & 31.241 & \bigO{n^{4.966}} & 14 & \(\bigO{n^{3.808}}\)
 \\
 \FMMA{2}{2}{2}{7} Str69 & {12} & \bigO{n^{\mathbf{3.585}}}& 6.829 & \bigO{n^{2.772}} & {17.889} & \bigO{n^{\mathbf{4.161}}} & 10.453 & \(\bigO{n^{3.386}}\)
 \\
 \FMMA{2}{2}{2}{7} DPS24 & 17.48& \bigO{n^{4.128}} & {5.966} & \bigO{n^{2.577}} & 27.705 & \bigO{n^{4.793}} & {10.008} & \bigO{n^{3.323}}
 \\
 \FMMA{4}{4}{4}{48} DPS25 & 288 & \bigO{n^{4.085}} & 38.163& \bigO{n^{2.628}} & 1032.0 & \bigO{n^{5.006}} & 139.906& \bigO{n^{3.565}}
 \\
 \FMMA{4}{4}{4}{48} Here & 148 & \bigO{n^{3.605}} & {25.456}& \bigO{n^{\mathbf{2.335}}} & 515 & \bigO{n^{4.504}} & {91.21} & \bigO{n^{\mathbf{3.256}}}
 \\
 \bottomrule
 \end{tabular}
 \caption{Growth factor and error bound function comparison depending on the choice of norms in the accuracy formula \(\pnorm{\widehat{\beta(\mat{A},\mat{B})}-\mat{C}} \leq f_{p,q}\qnorm{\mat{A}}\qnorm{\mat{B}}\ulp+\bbigO{\ulp^2}\) where~\({{f_{p,q}}=\bigO{n^{\log_k \gamma_{p,q}}}}\).}\label{tab:gfcomp}
 \end{center}
\end{table}

In norms~\({(p,q)=(\infty,\infty)}\) and~\({(p,q)=(2,\infty)}\)
(namely, when the norm on the input matrices is the max-norm),
Strassen's algorithm reaches the best accuracy bounds.
Now when the input norm is the~\(2\)-norm~\({(p,q)=(\infty,2)}\)
and~\({(p,q)=(2,2)}\), the variant with the best accuracy bound is
the one proposed here:
it improves on the author's accurate~\FMMA{2}{2}{2}{7}
schemes~\cite{jgd:2024:accurate,Dumas:2026:autoaccurate}
and on the initial~\FMMA{4}{4}{4}{48}
scheme~\cite{Dumas:2026:autoaccurate} (the latter on all considered norms).
\par
When moving from a~\FMMA{2}{2}{2}{7} scheme to a~\FMMA{4}{4}{4}{48}
scheme, the growth factors increase, but their error
bound~\({f_{p,q}}\) are governed by their error bound exponent:
the logarithm of this factor in base~\(2\) or~\(4\) respectively.
The second sub-column for each choice of norm displays this asymptotic
error exponent allowing to compare the accuracy bound of the recursive
algorithm regardless of the splitting.
The algorithm proposed in this note is the one with the least exponent
for~\({(p,q)=(\infty,2)}\),~(\({\log_{4}\gamma_{\infty,2}\approx2.335}\)),
improving even further on the most accurate~\FMMA{2}{2}{2}{7}
algorithm.
\par
Note that this new algorithm presented in~\Cref{ssec:newalgo}, was
selected by considering a weaker yet smoother expression
of~\(\gamma_{2,2}\):
\begin{equation}
 \gamma_2 = \sum_{i=1}^{r}
 \qsnorm{\row{\mat{L}}{i}}\qsnorm{\row{\mat{R}}{i}}\qsnorm{\row{\Transpose{\mat{P}}}{i}} \geq \gamma_{2,2}.
\end{equation}
\begin{remark}\label{rem:gamma2}
The~\FMMA{4}{4}{4}{48} algorithm of~\cite{Dumas:2025aa} reaches~\({\gamma_2 =\left({1+\sqrt{2}}\right)\left(48+16\sqrt{6}\right)}\).
The variant presented here in~\cref{sec:TheAlgorithm}, has the minimal
growth factor on this De~Groot orbit:~\({\gamma_2=\left({1+\sqrt{2}}\right)64}\).
\end{remark}
This last algorithm presents the best known accuracy order of
approximation for a sub-cubic algorithm, as shown
in~\cref{tab:gfcomp}.
We thus now turn to see how this theoretical improvement of the growth
factor actually impacts practical computations.
For this we first provide an associated straight-line program in the
next section.
\section{Straight-line program for the more accurate rational~\(\FMMA{4}{4}{4}{48}\)}\label{sec:NumericalSchemeAndComplexity}
We here give the (\plinopt~generated~\cite{jgd:2024:plinopt})
straight-line programs (\SLPs) of~\cref{lst:L,lst:R,lst:P}, obtained
from the~\({L,R,P}\) matrices presented in~\cref{alg:444accurate}.
\begin{lstlisting}[label=lst:L,style=slp,caption=\SLP for input linear
 forms defined by matrix L~of~\cref{alg:444accurate}]
x32:=A44-A24; x37:=A44+A24; x34:=A14+A34; x26:=A14-A34; x31:=A22-A42; x30:=A22+A42; x29:=A33-A13; x28:=A33+A13; x40:=A12+A13; x41:=A32+A33; x36:=A42+A43; x35:=A22+A23; x33:=A41+A21; l2:=A41-A21-x31; x27:=A31-A11; l9:=A31+A11-x28; x38:=x33-x37; x39:=x26+x27; l15:=x27-x29; l11:=x34+x28; l7:=x34-x28; l34:=l11-x41; l26:=x40-l34; l5:=x30+x33; l13:=x30-x33; l23:=x31+x32; l28:=x31-x32; l24:=x26+x29; l25:=x30+x37; l38:=l23-x35; x22:=l13-l23; l4:=x27+x29; l37:=l4-x41; x17:=l15-l5; x23:=l24-l25; l44:=x17-x23; l47:=x17+x23; x20:=l7-l28; x19:=l2-l9; x42:=l2+l28; x43:=l7-l9; x25:=x42+x43; x24:=x42-x43; x18:=l37+l34; l41:=l13-x35; l8:=x36-l41; x16:=l38+l41; l36:=x22+x18; l32:=x22-x18; l40:=x20+x23; l43:=x20-x23; x21:=l4-l11; l45:=x21+x16; l42:=x21-x16; l39:=x17-x19; l33:=x17+x19; l46:=x19-x20; l35:=x19+x20; l6:=x36+l38; l10:=x24-l32; l14:=x24-l42; l12:=l26+x18; l1:=l45+x25; l31:=l36-x25; x15:=x39-x38; x10:=x39+x38; l21:=x15-l45; l18:=x15-l36; l20:=x10+l32; l27:=x10-l42; x14:=x35*2; x13:=x36*2; x12:=x41*2; x11:=x40*2; l16:=x14+l33; l17:=x14+l40; l0:=x11+l40; l19:=x11-l33; l3:=l39+x13; l30:=l43-x13; l29:=l43+x12; l22:=l39+x12;
\end{lstlisting}
\begin{lstlisting}[label=lst:R,style=slp,caption=\SLP for input linear
 forms defined by matrix R~of~\cref{alg:444accurate}]
r19:=B22-B24; r12:=B24-B21; r0:=B22+B24; r26:=B22+B23; r24:=B44-B41; r25:=B44+B41; r2:=B12-B13; r9:=B12+B13; r14:=B12+B14; r3:=B31+B33; r30:=B31-B33; r6:=B33-B32; r8:=B31+B34; r10:=B11-B13; r18:=B41+B43; r21:=B44-B42; y12:=(B23-B13-B33-B43)/2; r34:=r26-y12; r36:=r18+y12; y14:=(B14+B34+B44-B24)/2; r41:=y14-r8; r42:=r14-y14; y18:=(B11+B31+B41-B21)/2; r37:=y18-r12; r4:=y14-r37; r32:=r10-y18; y21:=r42-r32; r15:=y21-r9; y17:=r32+r42; r5:=r2-y17; r20:=y12-r32; r13:=r41+y18; r31:=r36-y18; y19:=(B22-B12-B42-B32)/2; r38:=y19-r6; r23:=y12-r38; y10:=r41+r38; r16:=y10+r3; r45:=r21-y19; r1:=r45-y14; r11:=y19-r34; r27:=r42+y19; y11:=r45-r36; y15:=r45+r36; y16:=r34-r37; y20:=r34+r37; r22:=y20-r19; r28:=y15-r25; r29:=r0-y16; r7:=r24-y11; y13:=r41-r38; r17:=y13+r30; r43:=(y13+y16)/2; r40:=y16-r43; r33:=(y10+y20)/2; r39:=r33-y20; r35:=(y17+y15)/2; r47:=r35-y15; r46:=(y11+y21)/2; r44:=r46-y21;
\end{lstlisting}
\par\medskip
\begin{lstlisting}[label=lst:hadamard,style=slp, caption=Hadamard products]
p0=l0*r0; p1=l1*r1; p2=l2*r2; p3=l3*r3; p4=l4*r4; p5=l5*r5; p6=l6*r6; p7=l7*r7; p8=l8*r8; p9=l9*r9; p10=l10*r10; p11=l11*r11; p12=l12*r12; p13=l13*r13; p14=l14*r14; p15=l15*r15; p16=l16*r16; p17=l17*r17; p18=l18*r18; p19=l19*r19; p20=l20*r20; p21=l21*r21; p22=l22*r22; p23=l23*r23; p24=l24*r24; p25=l25*r25; p26=l26*r26; p27=l27*r27; p28=l28*r28; p29=l29*r29; p30=l30*r30; p31=l31*r31; p32=l32*r32; p33=l33*r33; p34=l34*r34; p35=l35*r35; p36=l36*r36; p37=l37*r37; p38=l38*r38; p39=l39*r39; p40=l40*r40; p41=l41*r41; p42=l42*r42; p43=l43*r43; p44=l44*r44; p45=l45*r45; p46=l46*r46; p47=l47*r47;
\end{lstlisting}
\par
\begin{lstlisting}[label=lst:P,style=slp,caption=\SLP for input linear
 forms defined by matrix P~of~\cref{alg:444accurate}]
z55:=p15-p24; z49:=p44+p15+p24; z54:=p5-p25; z48:=p47-p25-p5; z53:=p9+p7; z51:=p46+p9-p7; z50:=p28+p2; z52:=p35+p28-p2; z29:=p33-p16; z47:=p33+p19; z38:=p32+p10; z46:=p32+p20; z28:=p36+p18; z39:=p36-p31; z45:=p39+p3; z26:=p39+p22; z43:=p40-p0; z36:=p40+p17; z34:=p42+p14; z41:=p42+p27; z31:=p43+p29; z40:=p43+p30; z33:=p45-p1; z30:=p45+p21; z19:=z29-z36; z12:=z29+z36; z42:=z54+z53; z25:=z54-z53; z35:=z55+z50; z44:=z55-z50; z32:=z52+z49; z37:=z52-z49; z27:=z51+z48; z63:=z51-z48; z24:=z39+z38; z59:=z39-z38; z13:=z47+z43; z23:=z47-z43; z14:=z31+z26; z22:=z31-z26; z20:=z41+z30; z60:=z41-z30; z11:=z45+z40; z10:=z45-z40; z62:=z37+z25; z61:=z37-z25; z17:=z34+z33; z15:=z34-z33; z57:=z35+z27; z16:=z35-z27; z56:=z46+z28; z58:=z46-z28; z18:=z42+z32; z21:=z42-z32; z64:=z63-z44; z65:=z63+z44; C12:=(z23-z20-z62)/4; C14:=(z15-z13-z57)/4; C22:=(z60-z57-z12)/4; C24:=(z19-z17-z62)/4; C31:=(z24-z16+z14)/4; C33:=(z61+z58-z22)/4; C41:=(z10+z61-z59)/4; C43:=(z11+z56-z16)/4; C13:=(z18-z58-z23)/4+p26+p34; C32:=(z20+z22+z21)/4-p11-p34; C44:=(z17-z10-z21)/4+p8-p41; C21:=(z59-z19-z18)/4-p13+p41; C34:=(z15+z14-z64)/4+p4+p37; C11:=(z24+z65-z13)/4-p12-p37; C23:=(z56-z65-z12)/4+p23+p38; C42:=(z11+z60+z64)/4-p6-p38;
\end{lstlisting}
These straight-line programs require:
\begin{itemize}
\item \(80\) additions and~\(4\) binary shifts (multiplications by~\(2\)) for \texttt{L};
\item \(68\) additions and~\(8\) binary shifts (divisions by~\(2\)) for \texttt{R};
\item \(108\) additions,~\(16\) binary shifts (divisions by~\(4\)) for \texttt{P}.
\end{itemize}
This gives a total of~\(284\) operations and a theoretical complexity bound of:
\begin{equation}
	\left(1+\frac{284}{48-16}\right)n^{2+\log_{4}\!{3}}-\left(\frac{284}{48-16}\right)n^{2}
\approx{9.875\,n^{2.792481250} - 8.875\,n^{2}}.
\end{equation}
From this, we give in~\cref{sec:AlternativeBasis} an alternative basis
variant with a constant factor of the complexity bound reduced to
only~\(7\) and whose accuracy remains similar.
\section{Alternative bases algorithm}\label{sec:AlternativeBasis}
Following~\cite{Karstadt:2017aa,Dumas:2025aa}, we present in this section
the~\({L,R,P}\) matrices of an alternative basis derived from the
\LRPRepresentation{L}{R}{P} matrices of~\cref{alg:444accurate}.
They satisfy that~\({{L=L_{\textup{alt}}\cdot{}L_{\textup{cob}}},
  {R=R_{\textup{alt}}\cdot{}R_{\textup{cob}}}}\)
and~\({P=P_{\textup{cob}}\cdot{}P_{\textup{alt}}}\), with a common
inner dimension of~\(47\).

Looking at the \SLP of \cref{lst:L}, we see, for instance, that
\texttt{l47:=l40-l43+l44}
(and that \texttt{l47} is not reused in this \SLP).
Therefore, in~\(0\)-based numbering, this means that row~\(47\) is the
linear combination of rows~\(40\), \(43\) and~\(44\).
Let~\(\mat{L}_{h}\) be the first~\(47\) rows of~\(\mat{L}\)
and~\(l_{47}\) the~\(48\)-th of~\cref{alg:444accurate} (top, left).
Let also~\(\vec{e_i}\) (again~\(0\)-indexed) be the~\(i\)-th vector of
the canonical basis.
We obtain the decomposition:
\begin{equation}
  \mat{L}=\begin{smatrix} \mat{L}_h \\ l_{47}\end{smatrix}=
  \mat{L}_{\textup{alt}} \cdot \mat{L}_{\textup{cob}} =
  \begin{smatrix}\mat{I}_{47}\\\vec{e_{40}}-\vec{e_{43}}+\vec{e_{44}}\end{smatrix}\cdot\mat{L}_h.
\end{equation}

Similarly, we see in~\cref{lst:R}, that, e.g.,
\texttt{r47:=r42+r44-r45}
(and that \texttt{r47} is not reused in this \SLP).
We can thus define
accordingly~\({\mat{R}_{\textup{alt}}=\begin{smatrix} \mat{I}_{47}
    \\ \vec{e_{42}}+\vec{e_{44}}-\vec{e_{45}}\end{smatrix}}\)
and~\({\mat{R}_{\textup{cob}}}\) as~\(\mat{R}\)
of~\cref{alg:444accurate} (top, right), with its last row removed.

Finally, looking at~\cref{alg:444accurate}, there is a linear
combination of~\(2\) columns of~\(\mat{P}\) that gives a third
one:~\({c_{0}=c_{17}-c_{40}}\).
We can therefore
define~\({\mat{P}_{\textup{alt}}=\Transpose{\begin{smatrix}
    \vec{e_{17}}-\vec{e_{40}}\\
    \mat{I}_{47}\end{smatrix}}}\)
and~\(\mat{P}_{\textup{cob}}\) as~\(\mat{P}\)
of~\cref{alg:444accurate}, bottom, with its first column removed.

The algorithm of these L, R, P matrices can be realized with
straight-line programs with, respectively,~\({2,2}\) and~\(2\)
additions (for the latter,~\(\Transpose{\mat{P}}_{\textup{alt}}\) can
obviously be realized with~\(1\) addition, and
therefore~\(\mat{P}_{\textup{alt}}\) requires~\({1+(48-47)=2}\)
additions, by the transposition principle).
This gives a theoretical complexity bound of:
\begin{equation}\label{eq:altbasecomp}
	\left(1+\frac{2+2+2}{48-47}\right)n^{2+\log_{4}\!{3}}
	-\left(\frac{2+2+2}{48-47}\right)n^{\log_{4}\!{47}}
\approx 7n^{2.792481250}.
\end{equation}
The cost induced by the respective change-of-basis matrices~\({\mat{L}_{\textup{cob}}, \mat{R}_{\textup{cob}}}\) and~\(\mat{P}_{\textup{cob}}\) is~\({o\!\left(n^{2+\log_{4}\!{3}}\right)}\) for~\cref{eq:altbasecomp}~\cite{Beniamini:2019aa}.
The associated straight-line programs require indeed:
\begin{itemize}
\item \(79\) additions and~\(4\) multiplications (binary shifts) for \texttt{CoB\_L};
\item \(67\) additions and~\(8\) divisions (binary shifts) for \texttt{CoB\_R};
\item \(107\) additions and~\(16\) divisions (binary shifts) for \texttt{CoB\_P}.
\end{itemize}
This is a total of~\(281\) operations for an added complexity bound of~\({\left(\frac{281}{47-16}\right)\left(n^{\log_{4}\!{47}}-n^2\right)=o\!\left(n^{\log_4{\!48}}\right)}\).
\section{Experiments}\label{sec:bench}
\begin{figure}\centering
\begingroup
  \makeatletter
  \providecommand\color[2][]{%
    \GenericError{(gnuplot) \space\space\space\@spaces}{%
      Package color not loaded in conjunction with
      terminal option `colourtext'%
    }{See the gnuplot documentation for explanation.%
    }{Either use 'blacktext' in gnuplot or load the package
      color.sty in LaTeX.}%
    \renewcommand\color[2][]{}%
  }%
  \providecommand\includegraphics[2][]{%
    \GenericError{(gnuplot) \space\space\space\@spaces}{%
      Package graphicx or graphics not loaded%
    }{See the gnuplot documentation for explanation.%
    }{The gnuplot epslatex terminal needs graphicx.sty or graphics.sty.}%
    \renewcommand\includegraphics[2][]{}%
  }%
  \providecommand\rotatebox[2]{#2}%
  \@ifundefined{ifGPcolor}{%
    \newif\ifGPcolor
    \GPcolortrue
  }{}%
  \@ifundefined{ifGPblacktext}{%
    \newif\ifGPblacktext
    \GPblacktexttrue
  }{}%
  \let\gplgaddtomacro\g@addto@macro
  \gdef\gplbacktext{}%
  \gdef\gplfronttext{}%
  \makeatother
  \ifGPblacktext
    \def\colorrgb#1{}%
    \def\colorgray#1{}%
  \else
    \ifGPcolor
      \def\colorrgb#1{\color[rgb]{#1}}%
      \def\colorgray#1{\color[gray]{#1}}%
      \expandafter\def\csname LTw\endcsname{\color{white}}%
      \expandafter\def\csname LTb\endcsname{\color{black}}%
      \expandafter\def\csname LTa\endcsname{\color{black}}%
      \expandafter\def\csname LT0\endcsname{\color[rgb]{1,0,0}}%
      \expandafter\def\csname LT1\endcsname{\color[rgb]{0,1,0}}%
      \expandafter\def\csname LT2\endcsname{\color[rgb]{0,0,1}}%
      \expandafter\def\csname LT3\endcsname{\color[rgb]{1,0,1}}%
      \expandafter\def\csname LT4\endcsname{\color[rgb]{0,1,1}}%
      \expandafter\def\csname LT5\endcsname{\color[rgb]{1,1,0}}%
      \expandafter\def\csname LT6\endcsname{\color[rgb]{0,0,0}}%
      \expandafter\def\csname LT7\endcsname{\color[rgb]{1,0.3,0}}%
      \expandafter\def\csname LT8\endcsname{\color[rgb]{0.5,0.5,0.5}}%
    \else
      \def\colorrgb#1{\color{black}}%
      \def\colorgray#1{\color[gray]{#1}}%
      \expandafter\def\csname LTw\endcsname{\color{white}}%
      \expandafter\def\csname LTb\endcsname{\color{black}}%
      \expandafter\def\csname LTa\endcsname{\color{black}}%
      \expandafter\def\csname LT0\endcsname{\color{black}}%
      \expandafter\def\csname LT1\endcsname{\color{black}}%
      \expandafter\def\csname LT2\endcsname{\color{black}}%
      \expandafter\def\csname LT3\endcsname{\color{black}}%
      \expandafter\def\csname LT4\endcsname{\color{black}}%
      \expandafter\def\csname LT5\endcsname{\color{black}}%
      \expandafter\def\csname LT6\endcsname{\color{black}}%
      \expandafter\def\csname LT7\endcsname{\color{black}}%
      \expandafter\def\csname LT8\endcsname{\color{black}}%
    \fi
  \fi
    \setlength{\unitlength}{0.0500bp}%
    \ifx\gptboxheight\undefined%
      \newlength{\gptboxheight}%
      \newlength{\gptboxwidth}%
      \newsavebox{\gptboxtext}%
    \fi%
    \setlength{\fboxrule}{0.5pt}%
    \setlength{\fboxsep}{1pt}%
    \definecolor{tbcol}{rgb}{1,1,1}%
\begin{picture}(8640.00,5460.00)%
    \gplgaddtomacro\gplbacktext{%
      \csname LTb\endcsname
      \put(219,817){\makebox(0,0)[r]{\strut{}$10^{-14}$}}%
      \csname LTb\endcsname
      \put(219,1994){\makebox(0,0)[r]{\strut{}$10^{-13}$}}%
      \csname LTb\endcsname
      \put(219,3172){\makebox(0,0)[r]{\strut{}$10^{-12}$}}%
      \csname LTb\endcsname
      \put(219,4349){\makebox(0,0)[r]{\strut{}$10^{-11}$}}%
      \csname LTb\endcsname
      \put(387,174){\makebox(0,0){\strut{}$32$}}%
      \csname LTb\endcsname
      \put(2443,174){\makebox(0,0){\strut{}$64$}}%
      \csname LTb\endcsname
      \put(4500,174){\makebox(0,0){\strut{}$128$}}%
      \csname LTb\endcsname
      \put(6557,174){\makebox(0,0){\strut{}$256$}}%
      \csname LTb\endcsname
      \put(8614,174){\makebox(0,0){\strut{}$512$}}%
    }%
    \gplgaddtomacro\gplfronttext{%
      \csname LTb\endcsname
      \put(801,5059){\makebox(0,0)[l]{\strut{}\FMMA{2}{2}{2}{7} Winograd \cite{winograd:1977:complexite}}}%
      \csname LTb\endcsname
      \put(801,4787){\makebox(0,0)[l]{\strut{}\FMMA{2}{2}{2}{7} Strassen \cite{strassen:1969}}}%
      \csname LTb\endcsname
      \put(801,4515){\makebox(0,0)[l]{\strut{}\FMMA{4}{4}{4}{48} DPS25 \cite{Dumas:2025aa}}}%
      \csname LTb\endcsname
      \put(801,4242){\makebox(0,0)[l]{\strut{}\FMMA{2}{2}{2}{7} DPS24 \cite{jgd:2024:accurate}}}%
      \csname LTb\endcsname
      \put(801,3970){\makebox(0,0)[l]{\strut{}\FMMA{4}{4}{4}{48} \cref{alg:444accurate}}}%
      \csname LTb\endcsname
      \put(801,3698){\makebox(0,0)[l]{\strut{}\FMMA{4}{4}{4}{48}+Alt.\ Basis \cref{sec:AlternativeBasis}}}%
      \csname LTb\endcsname
      \put(801,3425){\makebox(0,0)[l]{\strut{}Conventional \(\bigO{n^3}\)}}%
      \csname LTb\endcsname
      \put(4456,0){\makebox(0,0){\strut{}Square matrix dimension}}%
    }%
    \gplbacktext
    \put(0,0){\includegraphics[width={432.00bp},height={273.00bp}]{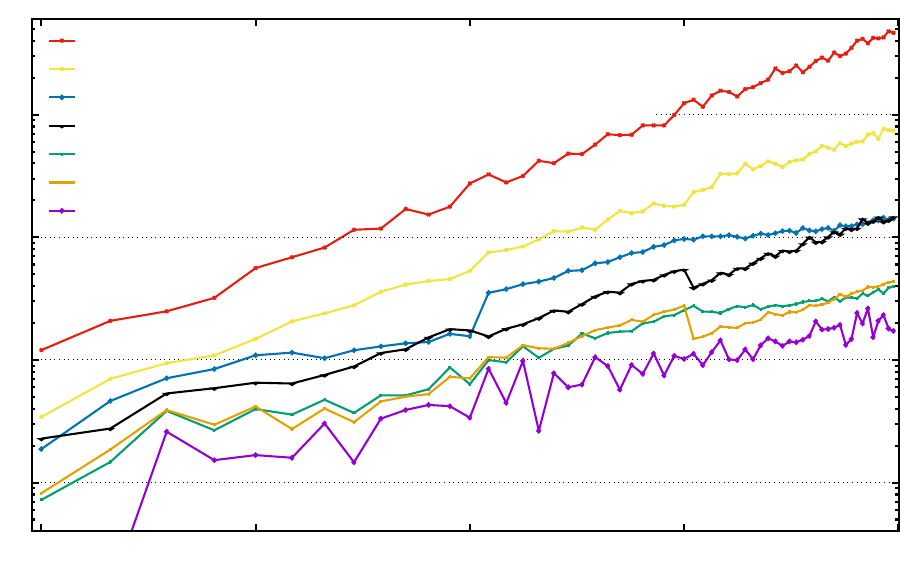}}%
    \gplfronttext
  \end{picture}%
\endgroup

 \caption{Numerical accuracy vs size (normal distribution)}\label{fig:accuracy}
\end{figure}

The accuracy obtained with our different fast variants is given in~\cref{fig:accuracy}.
For this, we use the Matlab framework of~\cite{jgd:2024:mFMM} and we present the error as the max-norm of the difference between the result of our implementations and the \emph{exact} matrix multiplication.
We see that the~\FMMA{4}{4}{4}{48} variants (\cite{Dumas:2025aa},~\cref{alg:444accurate,sec:AlternativeBasis}) are very close to the conventional cubic algorithm, and are more accurate than the~\FMMA{2}{2}{2}{7} ones (Winograd, Strassen,~\cite{Dumas:2026:autoaccurate}).
Note that the swift differences observed after the powers of two in size, are mostly due to the occurrence of an additional recursive level.
\par
Finally, note that these experiments display the error norm using the max-norm on the output~(\({p=\infty}\) in the notations of~\Cref{ssec:theoreticalbounds}) and regardless of any norm on the input (parameter~\(q\)).
The ranking by accuracy of these variants follows the one indicated by the second column of~\Cref{tab:gfcomp} (for~\({(p,q)=(\infty,2)}\)), the new algorithms of~\cref{alg:444accurate,sec:AlternativeBasis} being almost consistently more accurate.
\section{Concluding remarks}\label{sec:conclusion}
All the different matrices presented in this note can be found in the
\plinopt~library's data directory~\cite{jgd:2024:plinopt}.
For straight line programs presented
in~\cref{sec:NumericalSchemeAndComplexity} and associate matrices
in~\cref{sec:TheAlgorithm}, this is:
\begin{itemize}
\item\plinoptdata{data/4x4x4_48_accurate_{L,R,P}.s{ms,lp}};
\end{itemize}

The new variant proposed here stands, up to our knowledge, as the most accurate sub-cubic matrix multiplication algorithm currently known: namely it achieves the lowest exponent of asymptotic drift in accuracy considering the 2-norm on the input.
This is reflected in practice on experiments showing the sharpest accuracy on the normal distribution.
\bibliographystyle{plainurl}
\bibliography{RFC2603.bib}
\end{document}